\documentclass{aa}
\def\e{et~al.\ }
\def\es{{\rm erg\,s^{-1}} }

\newcommand{\be}{\begin{equation}}
\newcommand{\ee}{\end{equation}}
\newcommand{\bdm}{\begin{displaymath}}
\newcommand{\edm}{\end{displaymath}}

\begin{document}

   \title{Supersonic propeller spindown of neutron stars in
wind-fed mass-exchange close binaries}

   \author{N.R.\,Ikhsanov\inst{1,2}}


   \institute{Max-Planck-Institut f\"ur Radioastronomie, Auf dem
              H\"ugel 69, D-53121 Bonn, Germany\\
              \email{ikhsanov@mpifr-bonn.mpg.de}
              \and
              Central Astronomical Observatory of
              the Russian Academy of Sciences at Pulkovo,
              Pulkovo 65--1, 196140 Saint-Petersburg, Russia}

   \date{Received 19 November 2001 / Accepted 26 November 2001 }

\titlerunning{Supersonic propeller spindown of neutron stars}

    \abstract{
The supersonic propeller spindown of a neutron star moving in a
strong stellar wind of its massive companion is discussed. I show
that the supersonic propeller model presented by Davies \& Pringle
(\cite{dp81}) is self-consistent if the strength of the stellar
wind of the normal companion is $\dot{M}_{\rm c} \la 2.2\,10^{18}\
(M_{\rm ns}/M_{\sun})\ V_8\ {\rm g\,s^{-1}}$. Under these
conditions the model can be used for the interpretation of the
long-period pulsars in Be/X-ray transients. The spin history of
the neutron star in the long period Be/X-ray transient A0535+26 is
considered.
   }

   \maketitle


   \section{Introduction}

According to the present views on the evolution of interacting
close binary systems (e.g. Lipunov \cite{l92} and references
therein), a newly formed neutron star is presumed to rotate
rapidly, with a period of a fraction of a second. During the
further system evolution, the rotational rate of the neutron star
decreases, initially by the generation of the magneto-dipole waves
and ejection of relativistic particles ({\it pulsar-like
spindown}) and later by means of the interaction between the
magnetosphere of the neutron star and the stellar wind of its
companion ({\it propeller spindown}). When the spin period of the
neutron star reaches the critical value (so called break period,
$P_{\rm br}$) the accretion of material onto its surface begins
and the star switches on as an X-ray pulsar.

This scenario can be applied to the interpretation of the observed
periods of X-ray pulsars, provided the duration of the neutron
star spindown epoch $t_{\rm sd}$, is smaller than the time,
$t_{\rm ms}$, the companion star spends on the main sequence. In
the particular case of massive X-ray close binaries which display
pulses with long periods ($P_{\rm s} \ga 100$\,s), this condition
implies the spindown rate of neutron stars during the previous
epoch
  \bdm
\dot{P} \ga 2\,10^{-13}\ \left[\frac{P_{\rm s}}{100\,{\rm
s}}\right] \left[\frac{t_{\rm ms}}{10^7\,{\rm yr}}\right]^{-1}.
   \edm
Since this value is essentially larger than the observed average
spindown rate of neutron stars (see e.g. Taylor \e \cite{tml93}),
the question about the mechanism which can be responsible for the
required spindown rate arises.

A detailed investigation of the spindown epoch of neutron stars in
massive wind-fed close binaries has been presented by Davies \e
(\cite{dfp79}) and Davies \& Pringle (\cite{dp81}, hereafter
DP81). In particular, they have shown that the star magnetosphere
during the propeller spindown epoch is surrounded by a hot,
spherical quasi-static envelope. The interaction between the
magnetosphere and the envelope leads to the deceleration of the
rotation rate of the neutron star: the rotational energy loss by
the star is convected up through the envelope by the turbulent
motions and lost through its outer boundary. The neutron star
remains in the propeller state as long as the energy input to the
envelope due to the propeller action by the star dominates the
radiative losses from the envelope plasma. According to DP81, this
condition is satisfied if (i)\,the spin period of the star is
smaller than the break period (a subsonic propeller case), and
(ii)\,the strength of the normal companion stellar wind,
$\dot{M}_{\rm c}$, is smaller than the critical value
$\dot{M}_{\rm max}$ (a supersonic propeller case).

The first item was discussed in my previous paper (Ikhsanov
\cite{i01a}). In this letter I address the second item. I show
that the value of $\dot{M}_{\rm max}$ is by a factor of $10^3$
larger than that estimated in DP81. After this correction, the
duration of the spindown epoch proves to be essentially reduced.
This fact significantly increases the number of long-period X-ray
pulsars which can be interpreted within the canonical spindown
scenario.

    \section{Evaluation of $\dot{M}_{\rm max}$}

I consider a close binary system consisting of a fast rotating,
magnetized neutron star and an O or B type main sequence companion
which underfills its Roche lobe and loses mass in the form of
stellar wind. The neutron star, which moves through the wind of
its companion, is assumed to be in the state of {\it supersonic
propeller}. This means that the magnetospheric radius of the
neutron star exceeds its corotation radius, but is smaller than
both the radius of the light cylinder and the accretion radius.
Under these conditions the magnetosphere of the neutron star is
surrounded by a turbulent spherical plasma atmosphere, in which
the plasma pressure $p \propto R^{-3/2}$ and the plasma density
$\rho \propto R^{-1/2}$. The temperature throughout the atmosphere
is of the order of the free-fall temperature,
   \be\label{tff}
T(R) \simeq T_{\rm ff}(R)=(GM_{\rm ns} m_{\rm p})/(k R),
   \ee
and the sound speed, $V_{\rm s}$, as well as the velocity of
turbulent motions, $V_{\rm t}$, are of the order of the free-fall
velocity
     \be\label{vff}
V_{\rm s} \sim V_{\rm t} \sim V_{\rm ff}(R)=\sqrt{2GM_{\rm ns}/R}.
    \ee
Here $m_{\rm p}$ is the proton mass, $k$ and $G$ are the Boltzmann
and the gravitational constants, respectively, and $M_{\rm ns}$ is
the mass of the neutron star. The Mach number throughout the
envelope is $M_{\rm Mach}\equiv V_{\rm t}/V_{\rm s} \simeq 1$ (see
for discussion DP81, page\,213).

The atmosphere is extended from the magnetospheric boundary up to
the accretion radius of the neutron star
    \be
R_{\alpha} \equiv (2 GM_{\rm ns})/V_{\rm rel}^2,
  \ee
where $V_{\rm rel}$ is the relative velocity between the neutron
star and the wind of the normal companion. The plasma pressure at
the outer edge of the atmosphere is equal to the ram pressure of
the surrounding gas, which overflows the atmosphere as the neutron
star moves through the stellar wind of its normal companion. The
mass overflow rate, which is usually called the {\it strength of
the stellar wind}, is
    \bdm
\dot{M}_{\rm c} = \pi R_{\alpha}^2 \rho_{\infty} V_{\rm rel},
   \edm
where $\rho_{\infty}$ is the plasma density of the stellar wind
just beyond the outer edge of the atmosphere.

As shown by Davies \& Pringle, the supersonic propeller model
remains self-consistent as long as the energy input to the
atmosphere due to the propeller action by the neutron star
dominates the radiative losses. This condition can be expressed in
terms of the convective efficiency parameter (see DP81, page~221)
as
  \be\label{gamma}
\Gamma = M_{\rm Mach}^2 \left[\frac{V_{\rm t} t_{\rm
br}}{R}\right] = \frac{V_{\rm t}^3 t_{\rm br}}{V_{\rm s}^2 R} \ga
1.
    \ee
Here $t_{\rm br}$ is the bremsstrahlung cooling time:
  \be\label{tbr}
t_{\rm br} = 6.3 \times 10^4 \left[\frac{T}{10^9\,{\rm
K}}\right]^{1/2} \left[\frac{n}{10^{11}\,{\rm
cm^{-3}}}\right]^{-1} {\rm s},
  \ee
and $n$ is the number density of the atmospheric plasma, which at
the outer radius of the atmosphere can be evaluated as
  \be\label{n}
n(R_{\alpha}) \simeq \frac{\dot{M}_{\rm c}}{\pi R_{\alpha}^2
V_{\rm rel} m_{\rm p}}.
  \ee
Taking into account that in the supersonic propeller case $V_{\rm
s} \sim V_{\rm t} \propto R^{-1/2}$, $\rho \propto R^{-1/2}$, and
$T \propto R^{-1}$, one finds $\Gamma \propto R^{-3/2}$. This
indicates that the cooling dominates first at the outer radius of
the atmosphere and thus the model presented above is consistent if
$\Gamma(R_{\alpha}) \ga 1$.

Combining Eqs.~(\ref{tff} - \ref{n}) I
find this condition to be satisfied if the strength of the normal
companion stellar wind is $\dot{M}_{\rm c} \la \dot{M}_{\rm max}$,
where
 \be
\dot{M}_{\rm max} = 2.2\,10^{18}\ m\ V_8\ {\rm g\,s^{-1}}.
  \ee
Comparing this result with Eq.~(4.9) in DP81, one can conclude
that the value of the upper limit to the strength of the stellar
wind derived by Davies \& Pringle is underestimated by a factor of
1000.

    \section{Discussion}

The normal companions of more than half of the presently known
long period X-ray pulsars are Be/Oe stars (see e.g. Liu \e
\cite{lph00}). The stellar wind of these stars is not homogeneous.
It consists of a high velocity low density component at high
latitudes, and a low velocity high density circumstellar disk at
the equatorial plane. Since $\dot{M}_{\rm c} \propto \rho V_{\rm
rel}^{-1}$ the strength of the stellar wind at the equatorial
plane is essentially larger than that at high latitudes. This
particular property plays the key role in the interpretation of
Be/X-ray transients: the powerful ($L_{\rm x} \sim 10^{36} \div
10^{38}\,\es$) X-ray outbursts observed in these systems are
associated with the interaction between the neutron star and the
circumstellar disk surrounding its Be companion (see e.g.
Negueruela \cite{n98} and references therein). In the frame of
this model, the strength of the stellar wind in the disk is
$\dot{M}_{\rm c} \sim 10^{16} - 10^{18}\,{\rm g\,s^{-1}}$ that is
almost three orders of magnitude larger than $\dot{M}_{\rm max}$
derived by DP81. On this basis, Be/X-ray binaries were excluded
from the list of systems in which the long periods of neutron
stars can be interpreted within the spindown scenario presented by
Davies \& Pringle.

However the situation becomes completely different after applying
the correction to the value of $\dot{M}_{\rm max}$ presented in
this letter. Then the model of supersonic propeller constructed by
Davies \& Pringle proves to be valid even if the neutron star is
situated in a strong stellar wind moving through the circumstellar
disk of its Be/Oe companion. Furthermore, the spindown time scale
of neutron stars situated in a strong stellar wind is
significantly smaller than that of neutron stars in a weak stellar
wind. That is why the number of long-period pulsars which can be
analyzed within the canonical spindown scenario essentially
increases.

As an illustration, I consider a particular example of one of the
best studied Be/X-ray transients A0535+26. This system consists of
a 15\,$M_{\sun}$ Be star and the magnetized neutron star ($\mu
\simeq 10^{31}\,\mu_{31}\,{\rm G\,cm^3}$) rotating with the period
$P \simeq 103\,P_{103}$\,s. The rotational axis of the normal
companion is almost parallel to the orbital axis of the system, so
the trajectory of the neutron star lies in the plane of the
circumstellar disk surrounding the Be star. The average value of
the relative velocity between the neutron star and the surrounding
material is $V_{\rm rel} \simeq 10^{7}\,V_{7}\,{\rm cm\,s^{-1}}$
and the average strength of the stellar wind is $\dot{M}_{\rm c}
\simeq 10^{17}\,\dot{M}_{17}{\rm g\,s^{-1}}$ (see for discussion
Ikhsanov \cite{i01b} and references therein).

Under these conditions, $\dot{M}_{\rm max} \simeq 2\,10^{17}\,{\rm
g\,s^{-1}}$ is larger than $\dot{M}_{\rm c}$ and hence the
supersonic propeller model of DP81 can be used. Following this
model, the duration of the spindown epoch of the neutron star in
A0535+26 can be expressed as
  \be
\tau_{\rm sd} = \tau_{\rm a} + \tau_{\rm c} + \tau_{\rm d} \simeq
6\,10^6\ {\rm yr},
  \ee
where $\tau_{\rm a}$ is the time scale of the pulsar-like
spindown, which I evaluate following DP81 (see Eqs.~(3.3.5) and
(3.3.6) in their paper) as
   \be
\tau_{\rm a} = 4.8\,10^6\ \mu_{31}^{-3}\ \dot{M}_{17}^{-1/2}\
I_{45}\ V_7^{-1}\ {\rm s},
   \ee
$\tau_{\rm c}$ is the time scale of the supersonic propeller
spindown, which according to DP81 (see Eqs.~(3.1.7) and (3.3.6) in
their paper) is
  \be\label{tauc}
\tau_{\rm c} \simeq 10^6\ \mu_{31}^{-1}\ \dot{M}_{17}^{-1/2}\
I_{45}\ V_7^{-1}\ {\rm s},
  \ee
and $\tau_{\rm d}$ is the time scale of the subsonic propeller
spindown, which can be estimated according to Ikhsanov
(\cite{i01a}: Eq.~10) as
   \be\label{taud}
\tau_{\rm d} \simeq 10^3\ \mu_{31}^{-2}\ m\ I_{45}\ P_{103}\ {\rm
yr}.
  \ee
$I_{45}$ is the moment of inertia of the neutron star expressed in
units $10^{45}\,{\rm g\,cm^2}$.

The break period, at which the neutron star in A0535+26 switches
its state from subsonic propeller to accretor, can be evaluated
using Eq.~(8) in Ikhsanov (\cite{i01a}) as
  \be
P_{\rm br} \simeq 100\ \mu_{31}^{16/21}\ \dot{M}_{17}^{-5/7}\
m^{-4/21}\ {\rm s}.
  \ee
Thus, in the frame of the canonical spindown scenario, the neutron
star in A0535+26 is expected to decelerate its rotation, on the
time scale of the main-sequence lifetime of its companion, to the
presently observed spin period.

    \section{Conclusion}

Davies \& Pringle (\cite{dp81}) has underestimated the upper limit
to the strength of the stellar wind at which the supersonic
propeller model is self-consistent by a factor of $10^3$. The
incorporation of the re-estimated value into their spindown
scenario shows that the propeller mechanism can be responsible for
the origin of the long-period X-ray pulsars in Be/X-ray
transients. Application of this scenario to particular objects
will be presented in a forthcoming paper.

 \begin{acknowledgements}
I would like to thank Prof. J.\,Pringle for useful discussion. I
acknowledge the support of the Alexander von Humboldt Foundation
within the Long-term Cooperation Program.
\end{acknowledgements}

\end{document}